# Vision and Outlook: The Future of Particle Physics

**Ian Shipsey**
*University of Oxford*
*Oxford, United Kingdom*
*E-mail:* ian.shipsey@physics.ox.ac.uk

As a community, our goal is to understand the fundamental nature of energy, matter, space, and time, and to apply that knowledge to understand the birth, evolution and fate of the universe. Our scope is broad and we use many tools: accelerator, non-accelerator & cosmological observations all have a critical role to play. The progress we have made towards our goal, the tools we need to progress further, the opportunities we have for achieving "transformational or paradigm-altering" scientific advances: *great discoveries,* and the importance of being a united global field to make progress toward our goal are the topics of this talk.

Dedication

Gino Bolla

(September 25, 1968 - September 04, 2016)

A remarkable and unique physicist, friend and colleague to many.

## Introduction

ICHEP was like a tidal wave. A record 1600 abstracts were submitted, of which 600 were selected for parallel presentations and 500 for posters by 65 conveners. During three days of plenary sessions, 36 speakers from around the world overviewed results presented at the parallel and poster sessions. The power that drives the wave is the work of >20,000 colleagues around the globe: students, post docs, engineers, technicians, scientists and professors, toiling day and night, making the measurements and the calculations; sharing a common vision that this remarkable cosmos is knowable. Most could not come to Chicago but they are with us in sprit. They are responsible for the remarkable results we have seen this past week.

As a community, our goal is to understand the fundamental nature of energy, matter, space, and time, and to apply that knowledge to understand the birth, evolution and fate of the universe. Our scope is broad and we use many tools: accelerator, non-accelerator & cosmological observations all have a critical role to play. The progress we have made towards our goal, the tools we need to progress further, the opportunities we have for achieving "transformational paradigm-altering" scientific advances: *great discoveries,* and the importance of being a united global field to make progress toward our goal are the topics of this talk.

### 1.1 The Standard Model - be proud!

It is important to be proud of what our community has accomplished. Our history is a remarkable track record of success. Recounting it establishes our credibility with new generations of politicians and new generations of the public, and gives us the confidence to be worthy descendants of our forebears as we lay out ambitious plans for the future.

From the discovery of the electron in 1896, the nucleus in 1911 and the neutron in 1931- the particles that compose the atom- to the discovery of the Higgs boson in 2012 that enables atoms to exist, building an understanding of the universe has been a century in the making. Our community has revolutionized human understanding of the Universe and its underlying code, structure and evolution. Through careful measurement, observation and deduction we have developed remarkably successful prevailing theories - the Standard Models of particle physics and cosmology - that are highly predictive and have been rigorously tested, in some cases to one part in ten billion. These are among the highest intellectual achievements in the history of our species; they will be part of our legacy to future generations for eternity.

The potential now exists to revolutionize our knowledge again. The sense of mystery has never been more acute in our field. At the heart of the Standard Model is a Higgs field and particle that is an enigma. Dark matter holds our universe together but it challenges our understanding. For every gram of ordinary matter in the universe the other four are dark. We don't know the dark quantum, yet the evidence is overwhelming: galactic rotation curves, hot gas in clusters, the Bullet Cluster, Big Bang Nucleosynthesis, strong gravitational lensing, weak gravitational lensing, SN1a and the Cosmic Microwave Background. Dark energy drives our universe apart and accounts for 70% of the mass-energy of the universe. There are three distinct lines of evidence for dark energy: the SN1a Hubble Plot, Baryon Acoustic Oscillations in the distribution of galaxies, and the Cosmic Microwave Background. Dark energy is a mystery. What we know, and what we are made of, described by the Standard Model of particle physics, is just the tip of the iceberg.

The list of questions continues: how did matter survive the big bang? Some phenomena must have produced a small asymmetry between matter and antimatter so that when every billion matter particles annihilated a billion anti-matter particles, a single matter particle remained. We can trace our existence back to these survivors. Why are there so many types of particle? Why do the particles have such a large range of masses? Why does the pattern of particles (generations) repeat three times? Why do neutrinos have mass at all (in the Standard Model they are massless)? What powered cosmic inflation, the exponential expansion of the early universe by a factor of at least $10^{78}$ in volume, starting about $10^{-36}$s after the big bang until $10^{-33}$ -$10^{-32}$ s. It may be due to a scalar field - particle physics at the Planck scale. Following the inflationary period, the universe continued to expand at a slower rate, until dark energy became important.

In every area of our enquiry we have profound questions, and the number of questions increased with the discovery of the Higgs. Before the Higgs was found there was one central question: "*does the Higgs exist?*" After the Higgs discovery there are many more questions. What is the relation between the Higgs boson and Electro-Weak Symmetry Breaking (EWSB)? Is the mass of the Higgs natural or fine-tuned? If it is natural: what new physics/symmetry is operating to maintain its low mass? Does the Higgs we have found regularize the divergent $V_LV_L$ cross-section at high $M(V_LV_L)$ or is there new dynamics? Is the Higgs elementary or composite? Is the Higgs alone or are there other Higgs bosons? What is origin of the coupling to fermions? Does the Higgs couple to dark matter? Do Higgs decays violate CP? What is the relationship between the Higgs field and the cosmological electroweak phase transition?

Not only are there many new questions, but we are also at a very different place when we ask them today than we were before. From 1967 until 2012, the Standard Model guided research. It was good guide! We were aided by no-lose theorems for the W and the top, and to regularise the WW cross section either a new particle must exist (the Higgs), or there was new physics. Now the Standard Model is complete we know of no further no-lose theorems. In principle the Standard Model could be valid to the Planck scale.

It is worth considering just how dramatically different the situation with a roadmap and without a roadmap is. Perception (our understanding of the universe) is a dynamic combination of top-down (theory) and bottom-up (data driven) processing. The need for detail (the quality and quantity of the data) depends on the distinctiveness of the object and the level of familiarity we have with it. When we know the characteristics and context of what to expect, as we did with the discovery of the W, top quark, and Higgs, a little data goes a long way (top-down dominates). A visual analog of this is to present a subject with the well-known image of Lincoln (Figure 1) where the subject is first told they are about to see the face of a very famous historical figure and then shown an image with a paucity of data, they readily recognize it as the face of Lincoln because of the prior information they have been given and their familiarity with Lincoln is high. In the second part of the experiment the subject has their ability to use top down processing abruptly taken away. In effect, the roadmap is removed. They are shown an image that they have seen before, but it is an image with which they have much less familiarity, and no information is given about the image. The image is Dali's "The Persistence of Memory" (Figure 2). Most subjects do not recognize the image, and this remains true as successively better quality images are presented to them. It is only when the full resolution original is presented to the subject that they recognize it. From 1967 to 2012 particle physics was in a situation very similar to recognizing the image of Lincoln. Since 2012 we are in a situation where we are trying to recognize a Dali masterpiece, with little information to guide us. Without a roadmap we are dependent on bottom up information: we are in a data driven era.

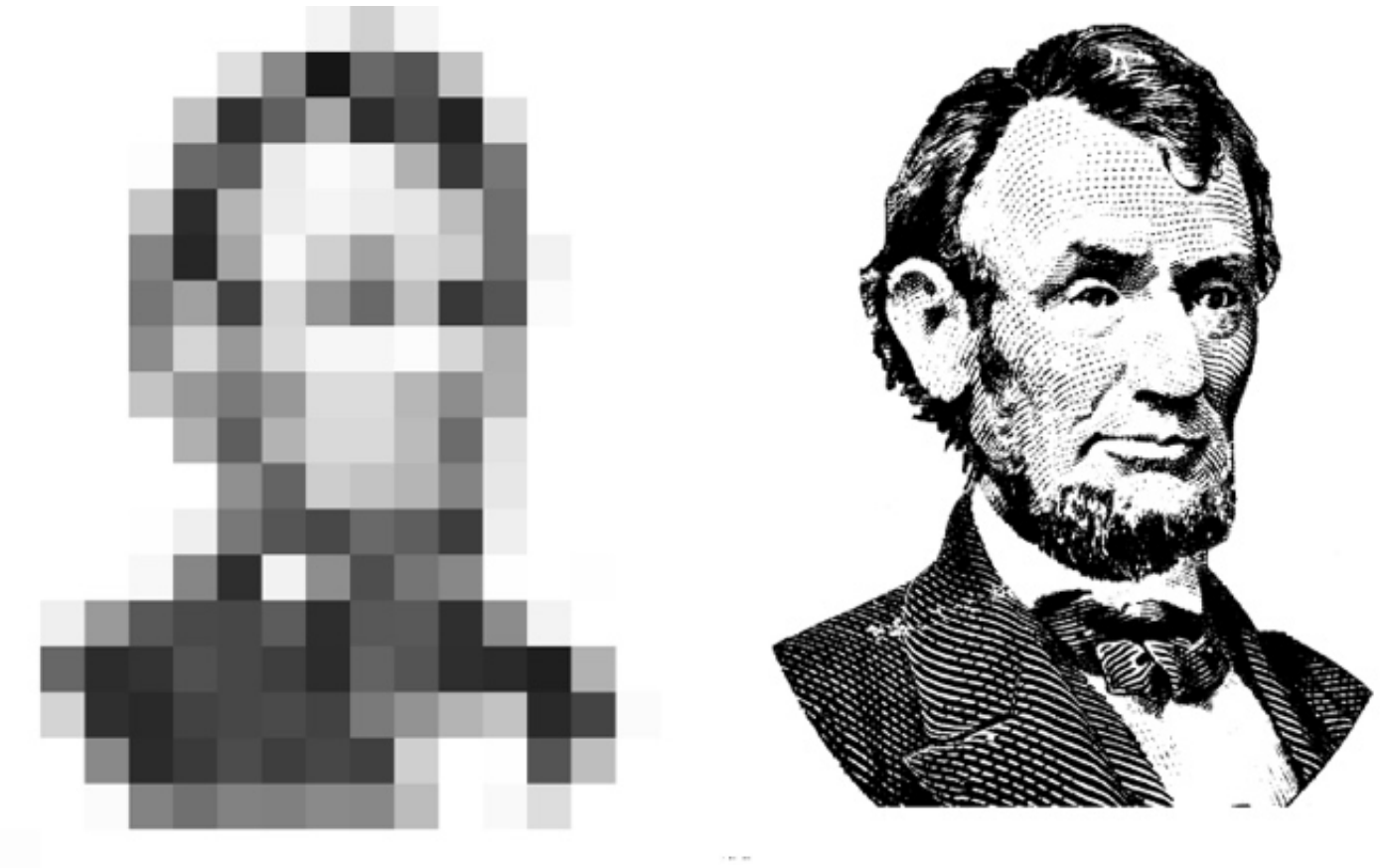

Figure 1: During the period 1967-2012 with the Standard Model as a guide top down processing dominated - a little data goes a long way. This is analogous to recognizing the Lincoln image (right) from the low-resolution image of Lincoln (left).

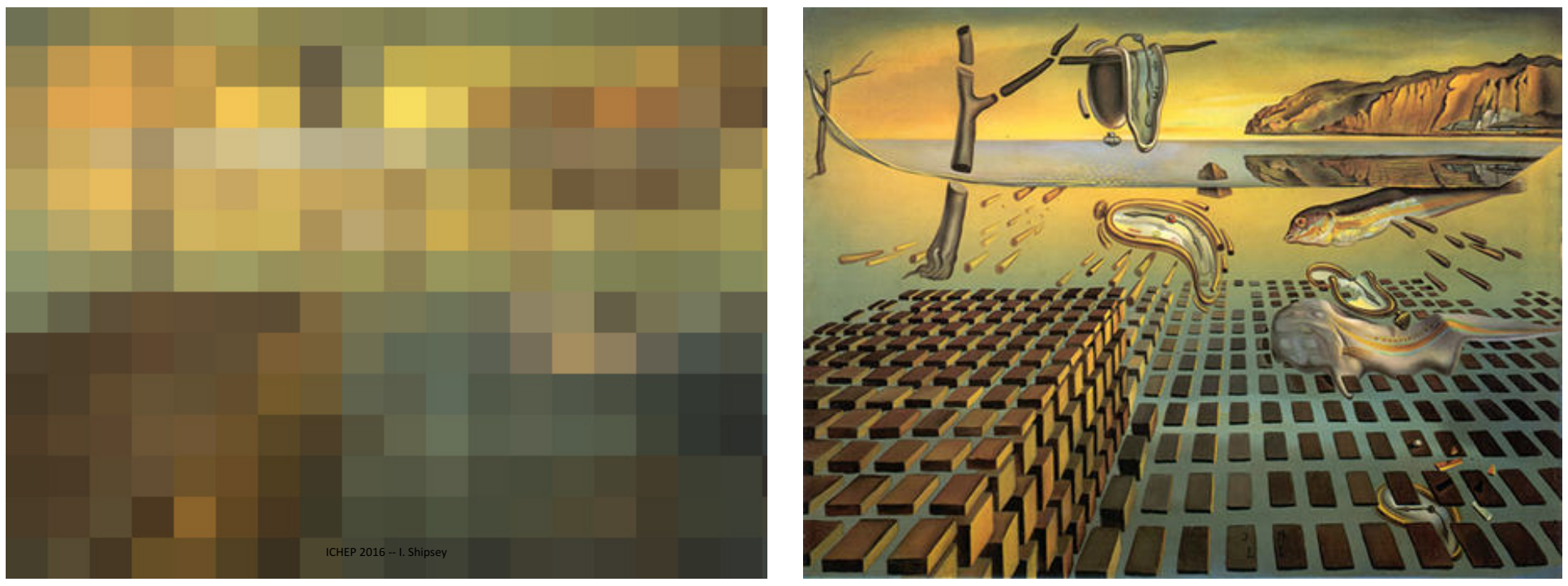

Figure 2: Today, there are no more no-lose theorems. Without a roadmap bottom-up dominates. This is analogous to recognizing the Dali masterpiece "The Disintegration of the Persistence of Memory" (right) from the low-resolution image of the same painting (left).

In consequence, our job is to, more than ever: "*Measure what is measureable and make measureable what is not so*." (Galileo). The discussion of the future of HEP must start from the understanding that there is no experiment or facility, proposed or conceivable, in the lab or in space, accelerator or non-accelerator, which can guarantee discoveries beyond the Standard Model and provide answers to the great questions of our field. We should not be surprised to find ourselves here, because the Standard Model is an effective theory. The Higgs is the first elementary spin 0 particle we have seen, it is the quantum of a field with an *ad hoc* potential, with a mass unprotected from quantum corrections that will act to increase it unless there is a remarkable degree of fine tuning. If the next scale is the Planck scale then this is what the degree of fine-tuning, in bare mass Planck units, looks like:

$$(M_H)^2 = 3.273{,}459{,}429{,}634{,}290{,}543{,}867{,}496{,}473{,}159{,}645$$
$$- 3.273{,}459{,}429{,}634{,}290{,}543{,}867{,}496{,}473{,}159{,}643$$

This seems to be an unnatural situation.

Some suggest that the universe is the way it is because, were it otherwise, we would not be here to observe it (the naturalness problem). The mass of the Higgs, the amount of dark energy (another perhaps fine-tuned quantity) and the values of other observables could be vacuum selection effects - our universe interpreted in terms of the multiverse. But it is premature to think so. As Alan Guth has said, *"it is the solution of last resort"*. As a community, we collectively choose to set the multiverse interpretation aside and seek to understand the universe. A balanced view of the naturalness problem was presented at ICHEP. While supersymmetry is still the leading explanation, theorists are also studying alternatives such as the "relaxion". This shifts attention to the dynamics of the early universe, with consequences that may be observable in future experiments.

Indeed, science progresses by experimentation, observation, and theory. Nobody would have predicted that slight irregularities in black body radiation would have led to an entirely new conception of the world in terms of quantum theory; that pondering the constancy of the speed of light would have led to $E= mc^2$; that special relativity and quantum mechanics would have led to anti-matter. Experiments that explore uncharted territory, or study phenomena we do not understand with greater precision, lead to a deeper understanding of nature. The global high-energy physics program does just that. Our work has the potential to lead to a reconciliation of the two great edifices of physics: Quantum Mechanics and General Relativity. We can be confident that our program will continue to reveal a cosmos more wonderful than we can possibly imagine.

This is not naive optimism. Not only do we already know of much we do not understand about the cosmos, but also our past experience points to one surprise after another. Often when we embark on an experiment with a goal in mind, we find something new and unexpected. That was the case for the following great discoveries: CP violation, the charm quark, the beauty quark, the top quark, the gluon, neutrino oscillations, and dark energy, to name just a few (see Figure 3). Precision instruments are a key to discovery when exploring new territory. Newton's statement still holds: "*what we know is a droplet what we don't know is an Ocean*"; the ocean is for us to explore. So at this unprecedented time in physics we need to be mindful of the words of Michelangelo, and aim big: *"The greater danger for most of us lies not in setting our aim too high and falling short; but in setting our aim too low, and achieving our mark"*. Is it OK to dream? It is a betrayal of humanity and generations that follow us to do otherwise. Working together is how we will make the case for our program, win funding for it, and enact it. Massive collaboration is the *modus operandi* of our field, and successful large international scientific collaborations are the proof. Our international collaborations inspire the public, made up of myriad individuals from across the globe, with diverse interests working together to achieve scientific goals. They seem to many to be an example of the best of humanity, and a model for how the world could be.

| Facility | Original purpose, Expert Opinion | Discovery with Precision Instrument |
|---|---|---|
| P.S. CERN (1960) | $\pi$ N interactions | Neutral Currents -> Z,W |
| AGS BNL (1960) | $\pi$ N interactions | Two kinds of neutrinos<br>Time reversal non-symmetry<br>charm quark |
| FNAL Batavia (1970) | Neutrino Physics | bottom quark<br>top quark |
| SLAC Spear (1970) | ep, QED | Partons, charm quark<br>tau lepton |
| ISR CERN (1980) | pp | Increasing pp cross section |
| PETRA DESY (1980) | top quark | Gluon |
| Super Kamiokande (2000) | Proton Decay | Neutrino oscillations |
| Telescopes (2000) | SN Cosmology | Curvature of the universe<br>Dark energy |

Figure 3: Discoveries in particle physics. Often when we embark on an experiment with a goal in mind we find something new. (From an original slide by S.C.C. Ting.)

To play a major role in this journey of discovery is the aspiration of our field and we should always remember that it is a rare privilege to participate in it. It is only possible to have come this far, and to go further, thanks to the taxpayers of our nations and the wisdom of governments who invest their money in science. Given the magnitude and breadth of the opportunities in front of us, the resources required to grasp them, and the global nature of our field, our long-term strategy has to maintain an international perspective, and strong international partnerships will be crucial to our future health. Crucial allies are governments and the public. There is an unprecedented interest in our field. Two prime examples are: 1) the LHC. The experiments and the observation of a Higgs boson that became a global phenomenon in 2012; and 2) the discovery of gravitational waves in 2016, that had even greater global appeal. These discoveries are opportunities to expand engagement with the public, our colleagues, and the governments of our nations, and to communicate what we have learned and the opportunities for discovery in particle physics: *the narrative of our field*.

We must explain to governments and the public of our nations why the world needs a healthy particle physics program. There are at least three reasons:

1) *Our science is important for our nations to pursue*. What is the world made of? What holds the world together? How did the world begin? For millennia all great societies have asked these questions.
2) *The big questions we ask attract young talent to all of the sciences*. The questions we ask are so big and so simple that almost everyone, from children to the general public, can understand and relate to them in some way. While many, many factors go into a decision to pursue a career in science, certainly one factor is the perception of big fundamental questions out there waiting to be answered. Our field helps to provide that perception in an important way. We help draw people to the physical sciences and help fill the education pipeline with talent.
3) *Particle physics is an essential part of the fabric of the physical sciences*. It contributes broadly to other physical sciences: (a) Accelerator science. The history of particle-physics-driven innovation in accelerator science is a resource for our nations. Some specific innovations include cable for pulsed superconducting magnets, the Klystron and the development of light sources. (b) Detector development, for example, Positron Emission Tomography. (c) Large Scale computing driven by large collaborations, for example, the World Wide Web. Of course it's a two-way street: other physical sciences contribute to particle physics as well.

### 1.2 What does a healthy particle physics program look like?

What does a healthy particle physics program look like? Some of the essential ingredients are a program focussed on the most compelling science, infrastructure to support the development of the tools required, and a long term vision and strategy to guide the program for future decades.

The field has historically been competitive between the regions; facilitating the regions to work together has been given a lot of thought. Here is a way we believe this can work: the regions can together address the full breadth of the field's most urgent scientific questions by each hosting unique world-class facilities at home, and partnering in high-priority facilities hosted elsewhere. Both hosting and partnering are an essential component of an achievable global vision. Strong foundations of international cooperation exist, with the Large Hadron Collider (LHC) at CERN serving as an exemplar of a successful large international science project. This model has been adopted by the LBNF/DUNE at FNAL.

Our big science with big tools requires the infrastructure to support the development of those tools for today and for the future. Intimately related to the need for healthy infrastructure is the need for a strong program in accelerator R&D. The future of particle physics at the energy and intensity/precision frontiers is dependent on innovations in accelerator science. Reliable partnerships are essential for the success of international projects. This global perspective has found worldwide resonance. The 2013 *European Strategy for Particle Physics* report focuses, at CERN, on the Large Hadron Collider (LHC) program and envisions substantial participation at facilities in other regions. Japan, following its 2012 *Report of the Subcommittee on Future Projects of High Energy Physics*, expresses interest in hosting the International Linear Collider (ILC), pursuing the Hyper-Kamiokande experiment, and collaborating on several other domestic and international projects. The 2014 U.S *P5 Report* highlights collaboration on the most important scientific opportunities wherever they are, and to host unique, world-class facilities that engage the global scientific community including DUNE and cosmic frontier experiments.

So how do we fund all of this? Politicians will say (i) What is the science case? Convince me that this project is scientifically excellent. (ii) What is the project plan? Convince me that you know what you are doing: that scope, costs and schedule are under control. (iii) What is the business case? Convince me that this is a good use of public money. As has been noted by others, we need three P's: a **P**ositive environment for science, **P**roject-specific benefits and **P**ersonal connections with policymakers.

For the first **P** we are in very good shape. The Higgs and gravitational wave discoveries have created an exceptionally positive environment. How about the second **P**: the project-specific benefits? Why do governments support science? It is not primarily to understand the universe. It is because of the technological innovations it spawns, and the skills needed. Science is an economic driver of job creation. A study has been conducted for the LHC using a cost-benefit analysis (CBA) methodology [1] widely used by governments and economists to evaluate the socio-economic impact of investment in projects. Until now, the application of CBA to research infrastructure (RI) has been hindered by claims that the unpredictability of future economic benefits of science creates a difficulty for any quantitative forecasts. It is certainly the case that the CBA of research infrastructure is complex, and that there is a risk of underestimation of benefits. Nevertheless, given the importance and the increasing cost of science, the potential advantages for decision-makers of exploring new ways to measure and compare social benefits and costs of large-scale research infrastructure cannot be exaggerated. For the LHC the authors conservatively estimated that there is around a 90% probability that benefits exceed costs, with an expected net present value of about 2.9 billion euro, not considering the unpredictable applications of scientific discovery. (Compare this to the estimated cost of the LHC which is ~ 6 billion euro.) Finally, how about the third **P?** Here there is much scope for improvement, as most practicing scientists, and even the most senior, often have limited interactions with politicians. As politicians themselves know: *"all politics is local"* (Speaker of the United States House of Representative Tip O'Neill the longest serving Speaker 1977-1987). Take every opportunity to develop deep connections to politicians. We must continually invite politicians to our institutions, and present to them the narrative of our field, then invite them back again and again. Building deep relationships with politicians is critical to the future of our field.

## 1.3 The Energy Frontier

The direct measurement of the Higgs boson is the key to understanding EWSB. The light Higgs boson must be explained. A program focussed on Higgs couplings to fermions and vector bosons to a precision of a few per cent or less is required to address this physics. This program starts with the LHC. The spectacular performance of the LHC during 2016, which saw about 20 $fb^{-1}$ of 13 TeV proton–proton collisions delivered to ATLAS and CMS by the time of the conference, gave both experiments unprecedented sensitivity to study the Higgs and look for new particles and interactions.

The discovery of the Higgs in 2012 was one of the most important in particle physics in the last thirty years. Already Run 2 of the LHC has produced more Higgs bosons than in Run 1, and the Higgs has been re-established in the new data with a significance of 10σ (Figure 4). The major focus of the new analyses is to determine the production of Higgs particles in association with a W or Z boson, or with a pair of top quarks and their decay patterns. These production and decay channels are important tests of Higgs properties, and so far the Higgs behaves just as the SM predicts, within the large uncertainties. The agreement is captured in Figure 5 (left), a beautiful verification of electroweak unification, and in Figure 5 (right), which demonstrates that the Higgs coupling is proportional to mass.

To within the precision of the measurements, the Higgs looks just like SM Higgs, but it is mandatory to verify this. To do so we need to measure the couplings precisely. Full exploitation of the LHC/HL-LHC is the path to a few percent precision in couplings and a 50 MeV precision in mass determination (Figure 6 left). Full exploitation of a precision electron positron collider (ILC, CLIC or circular $e^+e^-$ collider) is the path to a model-independent measurement of the Higgs width and sub-percent measurement of the couplings (Figure 6 right).

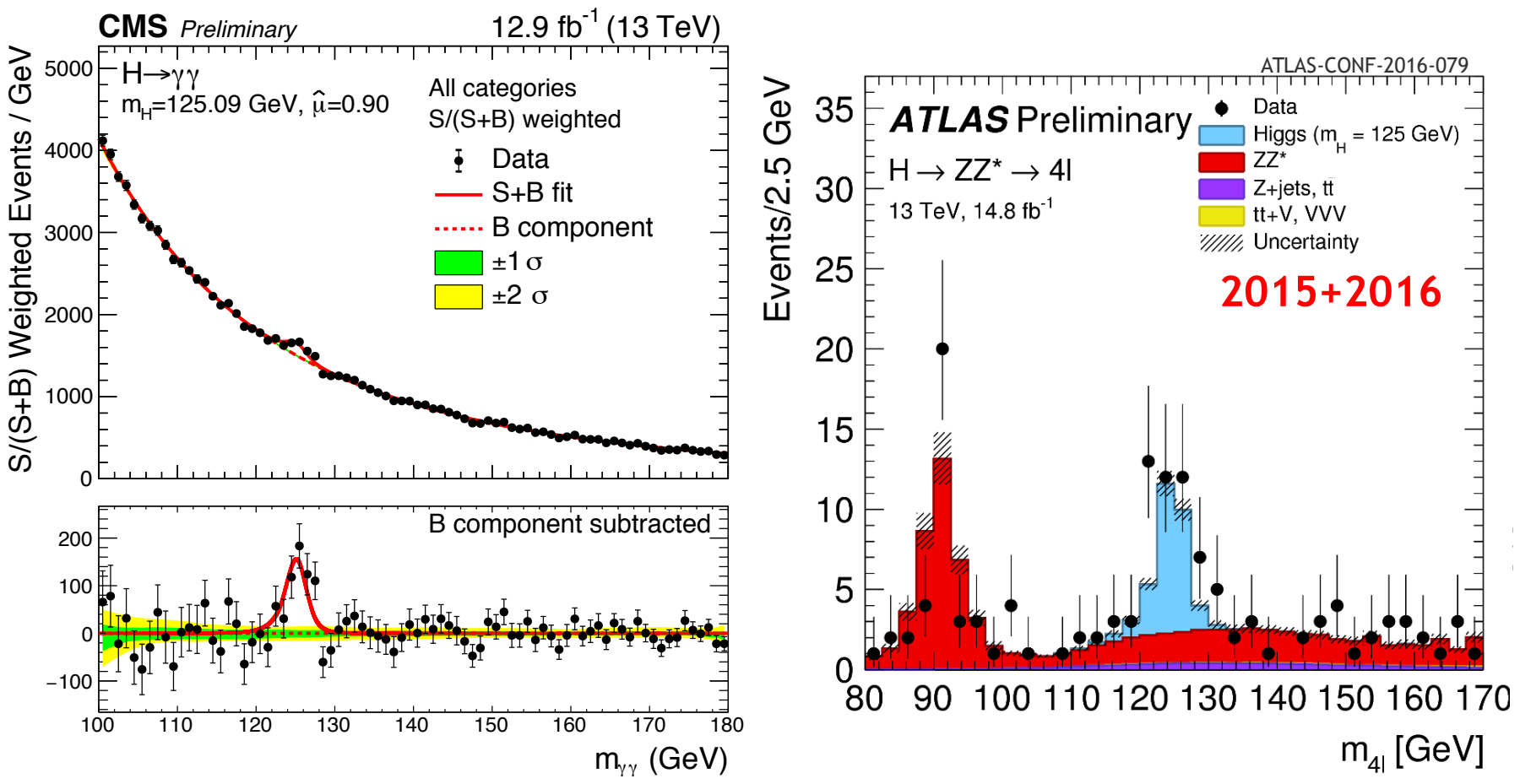

Figure 4: The CMS (left) and ATLAS (right) experiments rediscovery of the 125 GeV-mass Higgs Boson at almost twice the collision energy (13 TeV).

The Higgs must also have a dynamical property: it should be able to interact not only with other particles, but also with itself. What is the shape of the symmetry breaking potential and how is restored at high scales? The observable is the Higgs self-coupling cross section. It is difficult to measure due to the small cross section at LHC, but at a 100 TeV collider (FCC or a similar machine in China known as the CppC), with a much larger cross-section, it can be well-measured. The LHC will only probe the small quadratic oscillations around the symmetry breaking vacuum while a 100 TeV collider will give sensitivity to the functional form of the potential (Figure 7).

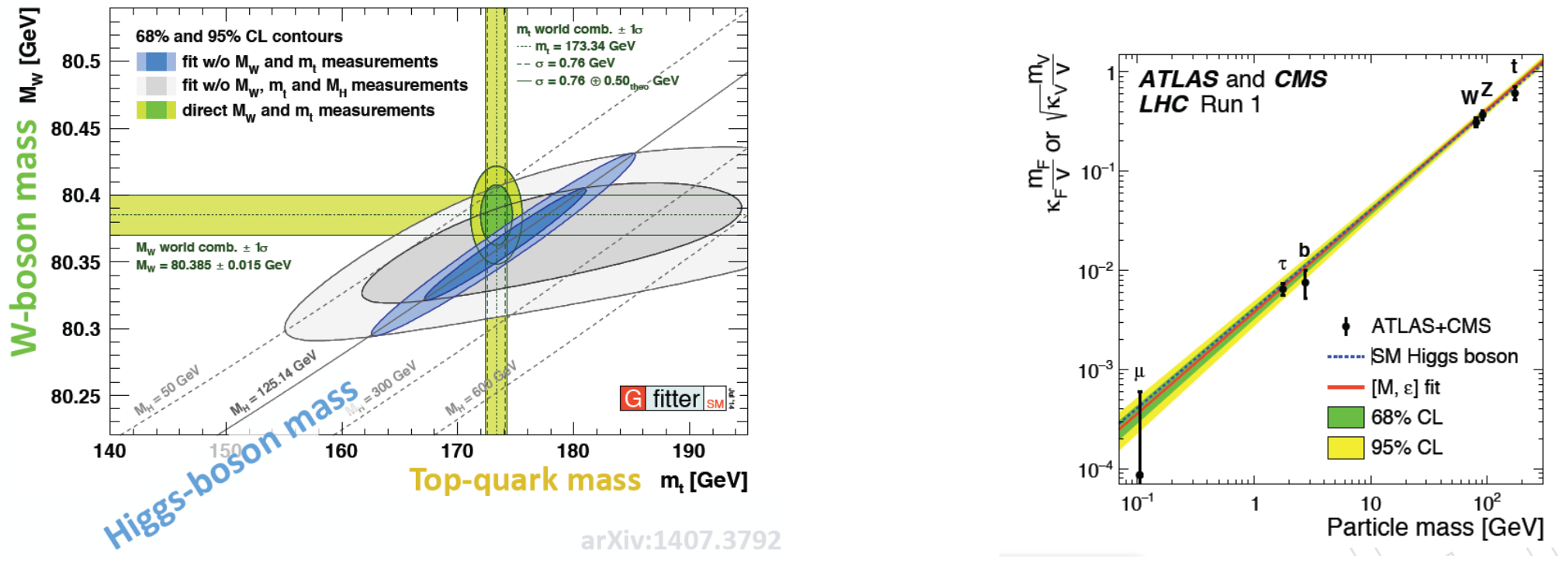

Figure 5: (left) the measured W mass compared to the measured top mass for given values of the Higgs mass compared to the prediction of the electroweak fit before the Higgs discovery (grey ellipse) and afterwards (blue ellipse). (Right) The Higgs coupling is proportional to mass.

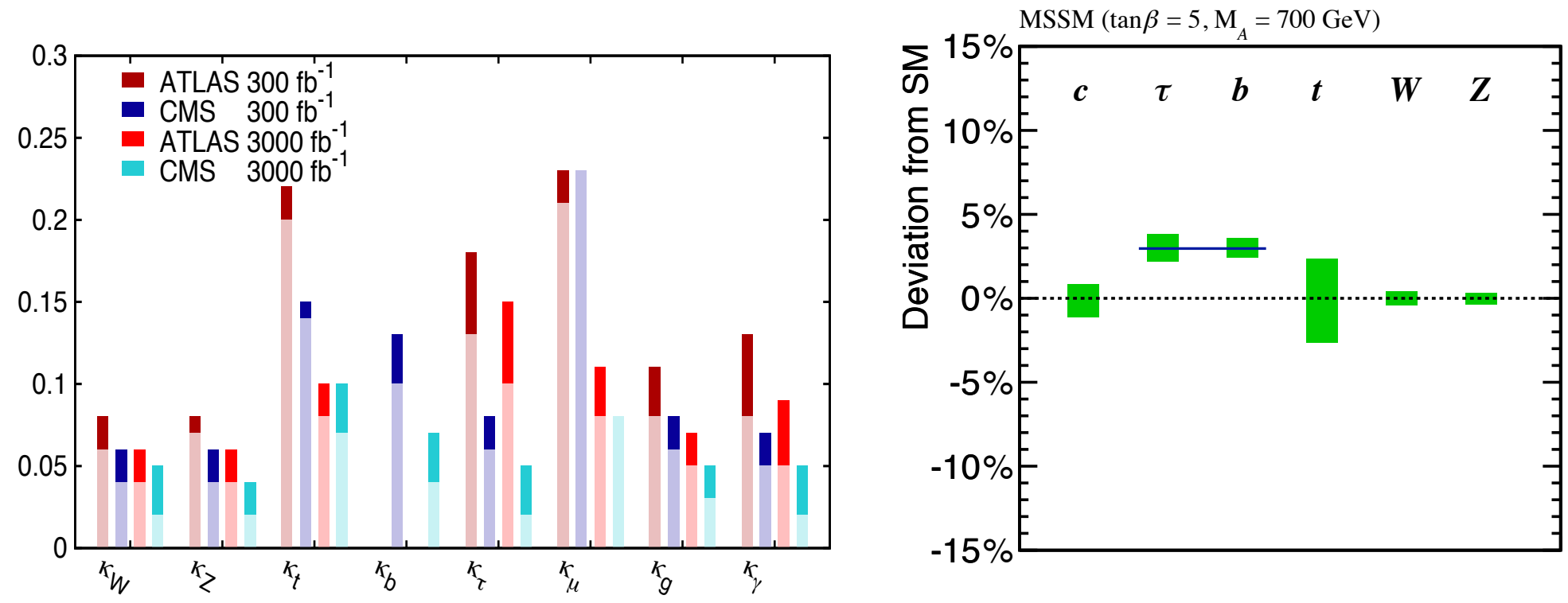

Figure 6: (left) Precision of Higgs couplings at ATLAS and CMS with 300 fb$^{-1}$ and 3,000 fb$^{-1}$. (Right) At the ILC the deviation from the Standard Model of the Higgs couplings in a representative MSSM with in green the uncertainty on the measurements.

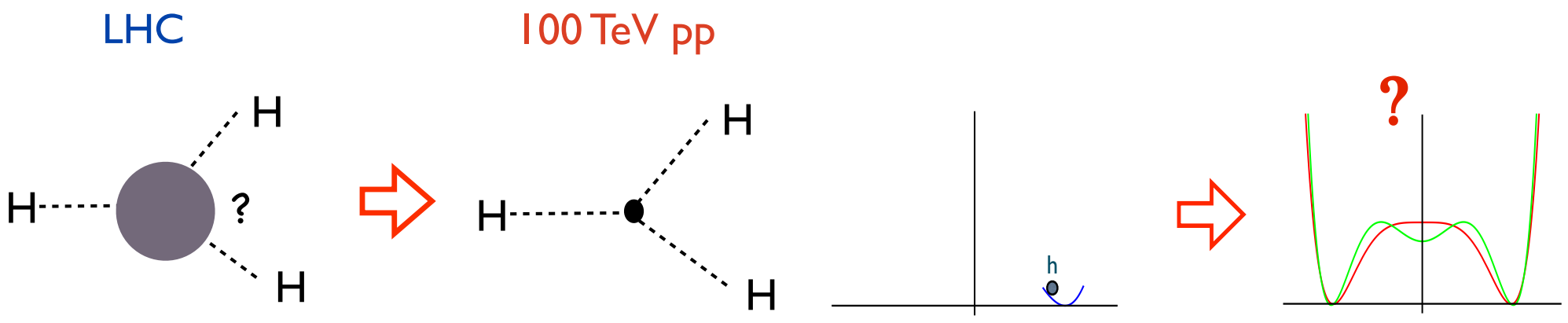

Figure 7: (Left) A sketch of one of the major advances obtained by going to a 100 TeV pp collider. The 100 TeV pp collider will see, for the first time, a fundamentally new dynamical process the self-interaction of an elementary particle uniquely associated with the Higgs. (Right) The LHC will only probe the small, quadratic oscillations around the symmetry breaking vacuum, without giving us any idea of the global structure of the potential [2].

Many new searches looking for heavier cousins of the Higgs were reported. These “heavy Higgs”, once produced, could decay in ways very similar to the Higgs itself, or might decay into a pair of Higgs bosons. Other searches cover the possibility that the Higgs boson itself has exotic decays: “invisible” decays into undetected particles, decays into exotic bosons or decays that violate the conservation of lepton flavour. No signals have emerged yet, but the LHC experiments are providing increasing sensitivity and coverage of the full menu of possibilities

TeV mass particles are needed in essentially all models of new physics. The search for them is imperative and integrally linked to searches for dark matter and rare processes. The most popular extension of the Standard Model is Supersymmetry (SUSY), the only unused symmetry of the Poincaré Group. There is a long list of reasons why SUSY is attractive. Among them: a solution to the hierarchy (naturalness) problem; a rationale for scalars; unification of the forces; a dark matter candidate; needed by string theory; helps the cosmological constant problem ($10^{-120}$ is reduced to $10^{-60}$). It is important to remember that SUSY is complex: it is *not* a single model but a large framework (Figure 8). SUSY is too big to explore without some assumptions**.** Searches in Run 2, are looking at more challenging scenarios than Run 1. Simplified models are explored at almost 2 TeV for gluons and almost a TeV for top squarks (Figure 8).

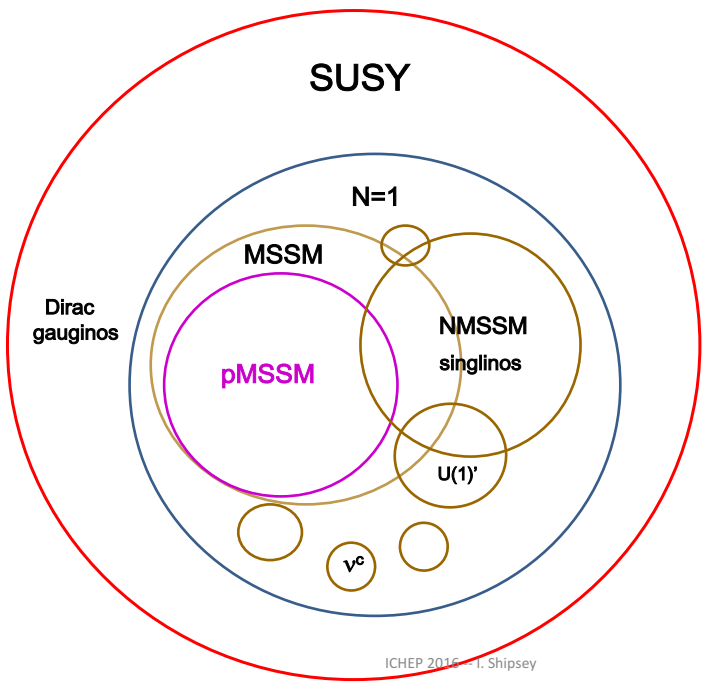

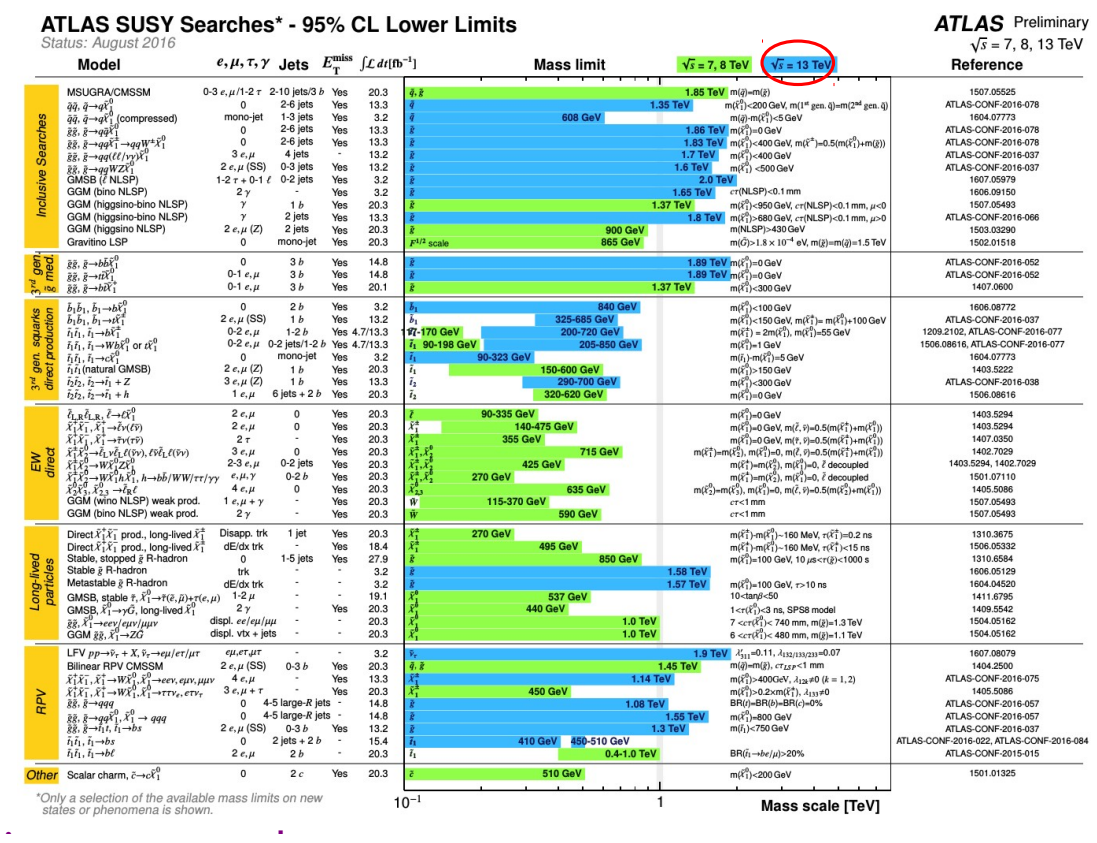

Figure 8: (Left) The SUSY framework showing, for example, the MSSM is a subset of SUSY. (Right) ATLAS SUSY Searches 95% CL lower limits from Run 1 and early Run 2 (preliminary). CMS presented similar results.

A 100 TeV collider will improve the reach for general direct searches by at least a factor of 5 (Figure 9). Taking natural SUSY as an example, stops and gluinos are light, but the first two generations may be heavier than 5 TeV. At 100 TeV, the reach for a heavy squark goes up to 35 TeV. The 100 TeV collider reach for neutralino dark matter is similarly impressive. While LHC can discover SUSY particles, to understand the type of SUSY we have found will take higher energies. At the moment, however, SUSY is not proving as simple to find, as was once thought it would be. The road to SUSY is fogbound, but the fog might clear at any moment. There are a huge range of models, both SUSY and non-SUSY, of physics beyond the Standard Model.

There is no doubt that many more that have not yet been thought of, and many viable directions that preserve naturalness are testable at the LHC, with a full elucidation of the physics to follow at 100 TeV.

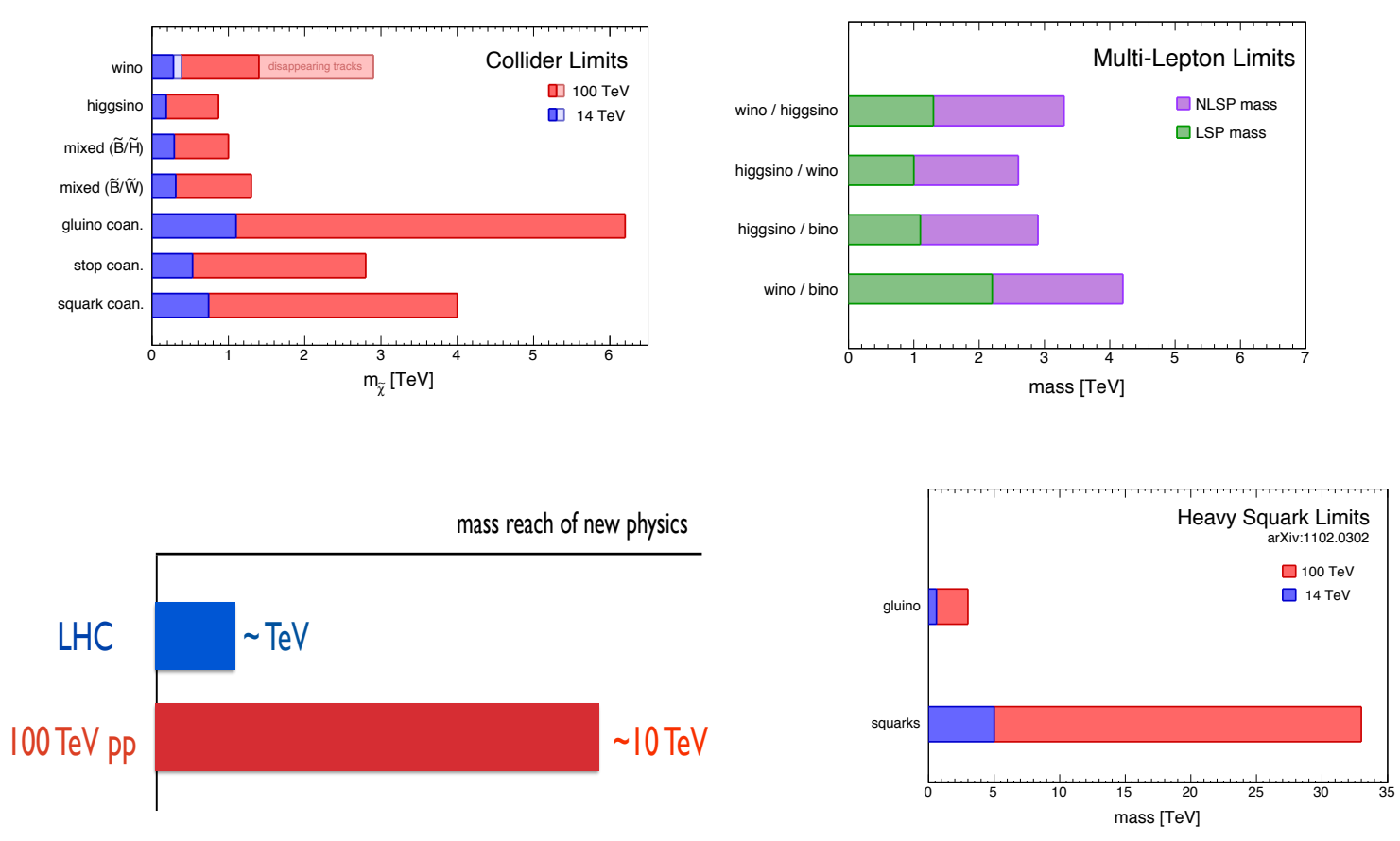

Figure 9: (Lower left) A sketch of one of the major advances obtained by going to a 100 TeV pp collider is an improvement in the reach of the direct search of new physics particles by at least a factor of 5. (Lower right) Reach for a heavy squark produced in association with a light gluino at a 100 TeV pp collider. (Upper left) reach for neutralino dark matter and electroweakino cascades. Composite figure from [2] and references therein.

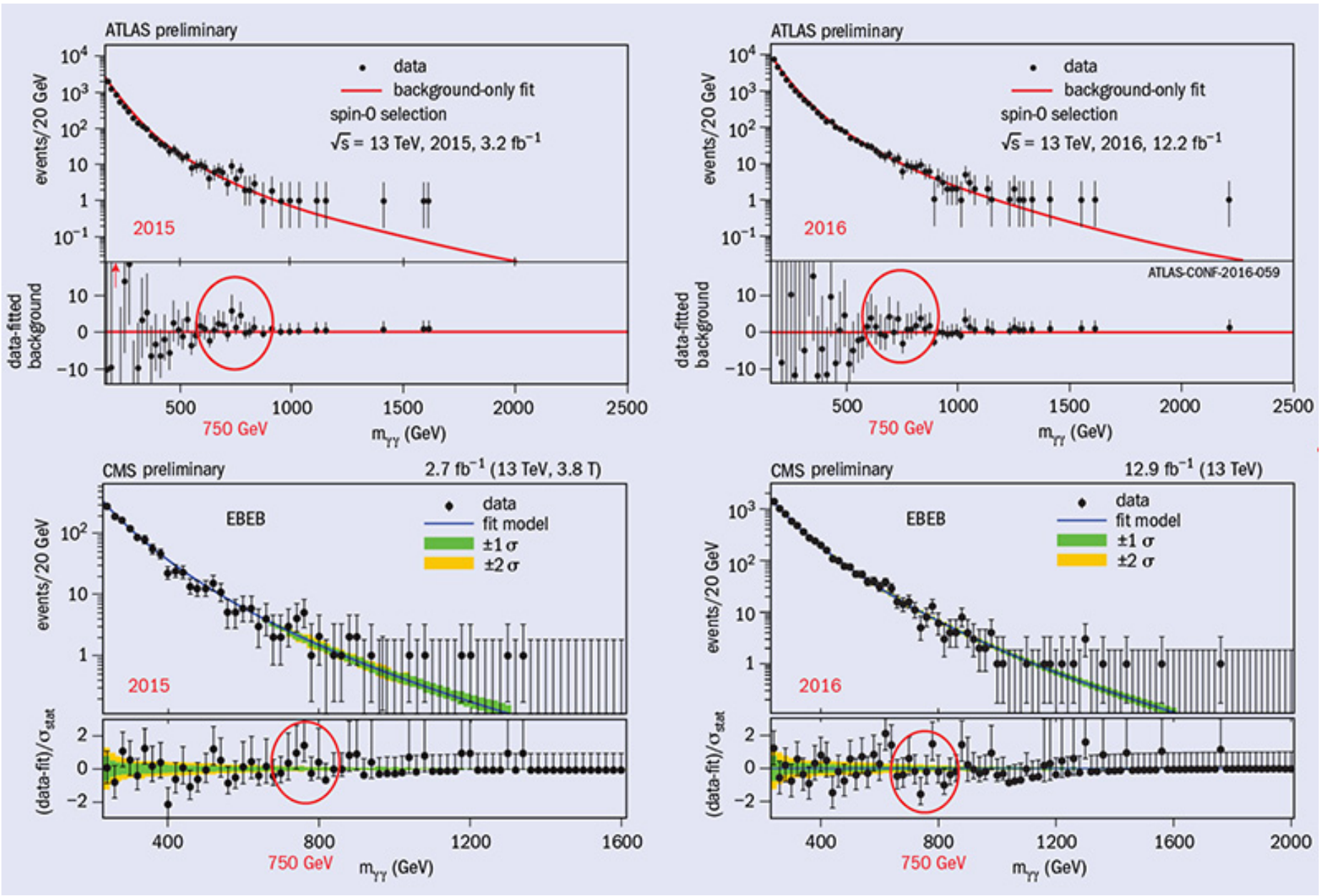

Figure 10: An excess in the number of diphoton events corresponding to a mass of 750 GeV observed by ATLAS and CMS in 2015 (left) did not reappear in data recorded in 2016 (right).

The collaborations reported on dozens of different searches for new phenomena at ICHEP. Both ATLAS and CMS revealed that their 2016 data do not confirm the previous hints of a di-photon resonance at 750 GeV (Figure 10). The number of preprints on the arxiv following the announcement last December is a testament to the fact we are in a data driven era, and appears to confirm Redman's theorem that *"any competent theoretician can fit any given theory to any given set of facts"*. Now that the attention span of the press is short, it seems to me we need to build a deep relationship, so that they understand our science is long term, needing investment over decades, and that we need their support as powerful allies in our relationship with the public. A case in point is the LHC, made possible by decades of investment in the CERN infrastructure, and the HL-LHC that will build on it.

Only ~2% of the complete LHC/ HL-LHC data set has been delivered as of summer 2016. There is every reason to be optimistic that an important discovery could come at any time. LHC Run 2 promises O(100) fb-1 by the end of 2018, typically a factor 10 in statistical power over Run 1 for measurements, and even more for searches. It is not far fetched to speculate that evidence for SUSY, or some other physics beyond the Standard Model, could be found in Run 2 and the headlines might look something like the cartoon in Figure 11.

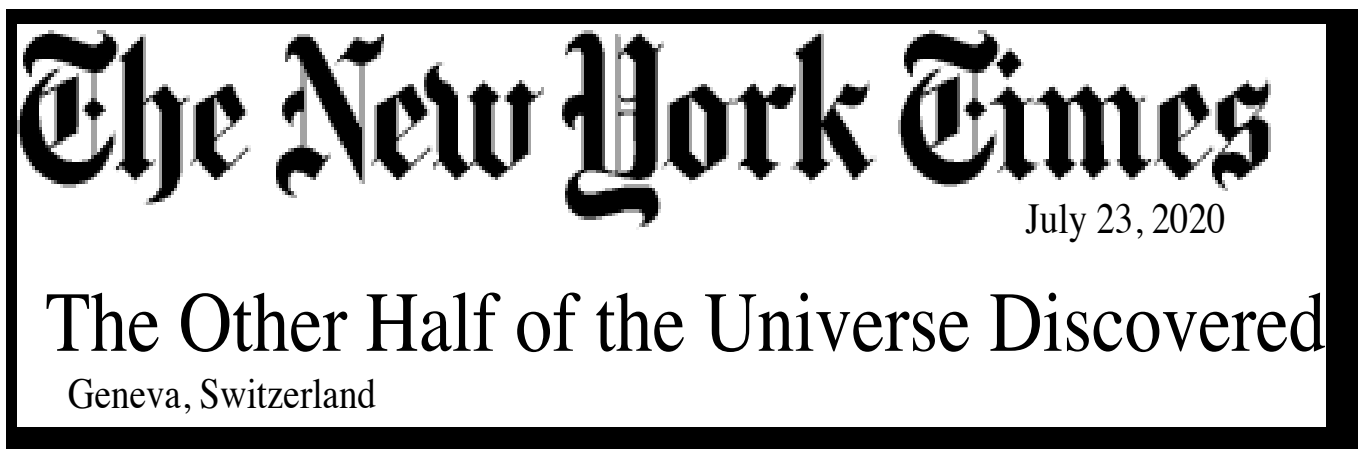
The New York Times
July 23, 2020
The Other Half of the Universe Discovered
Geneva, Switzerland

Figure 11: The announcement of the discovery of SUSY at the LHC in the New York Times in July, 2020. (Original cartoon from Hitoshi Murayama.)

### 1.4 The Intensity/precision frontier

New physics can show up at the intensity/precision frontier before the energy frontier, due to the mass reach of quantum loops. Ordinary beta-decay at MeV energies informs us of a virtual mediator at 80 GeV (the W). The GIM mechanism indicated the existence of charm, before the discovery of charm. The discovery of CP violation and its accommodation in the CKM matrix required the third generation before the discovery of beauty and top. The observation of neutral currents were evidence for the Z before the discovery of the Z.

Many intensity frontier experiments reported searches for new particles and interactions, including new LHCb results on the most sensitive search to date for CP violation in the decays of neutral D mesons, which, if detected, would allow researchers to probe CP violation in the up-type quark sector. The MEG (Mu to E Gamma) experiment at the Paul Scherrer Institute in Switzerland reported the most sensitive search to date for charged lepton-flavour violation, which would also be a clear signature of new physics. Using bottom and charm quarks to probe new physics, LHCb, the Beijing Spectrometer (BES) at IHEP in China and Belle at KEK in Japan presented a series of precision and rare-process results. While there are a few interesting discrepancies from Standard Model (SM) predictions, presently no signs of physics beyond the SM have emerged.

There is a need for more precision. Two quotes capture this nicely in the context of the discovery of CP violation in the Kaon system (remember: $B(K_L^0 \rightarrow \pi^+\pi^-) \sim 2 \times 10^{-3}$): *"Imagine if Fitch and Cronin had stopped at the 1% level, how much physics would have been missed"* – A.Soni. *"A special search at Dubna was carried out by Okonov and his group. They did not find a single $K_L^0 \rightarrow \pi^+\pi^-$ event among 600 decays into charged particles (Anikira et al., JETP 1962). At that stage the search was terminated by the administration of the lab. The group was unlucky."* – L.Okun [3].

Flavor physics at the LHC has been a great success, with run-1 delivering in all important topics, from discoveries ($B_s \rightarrow \mu\mu$), to great steps forward in knowledge of the CKM unitarity triangle angle $\gamma$ ($\phi_3$), and from precise studies of CP violation in the $B_s$ system to probing for CPV in charm with per mille precision. Some intriguing anomalies have emerged from LHC-b and the B-factories (see Figure 12). The quest for indirect discovery of new physics requires patterns of deviations to exist, and a pattern may be emerging in rare B decays to strange final states with dileptons. There are also hints of lepton universality violation in semileptonic B decays to charm. These modes are used in the determination of $V_{cb}$, where there are long-standing inconsistencies in the value of both $V_{cb}$ and $V_{ub}$ determined by the inclusive and exclusive semileptonic final states.

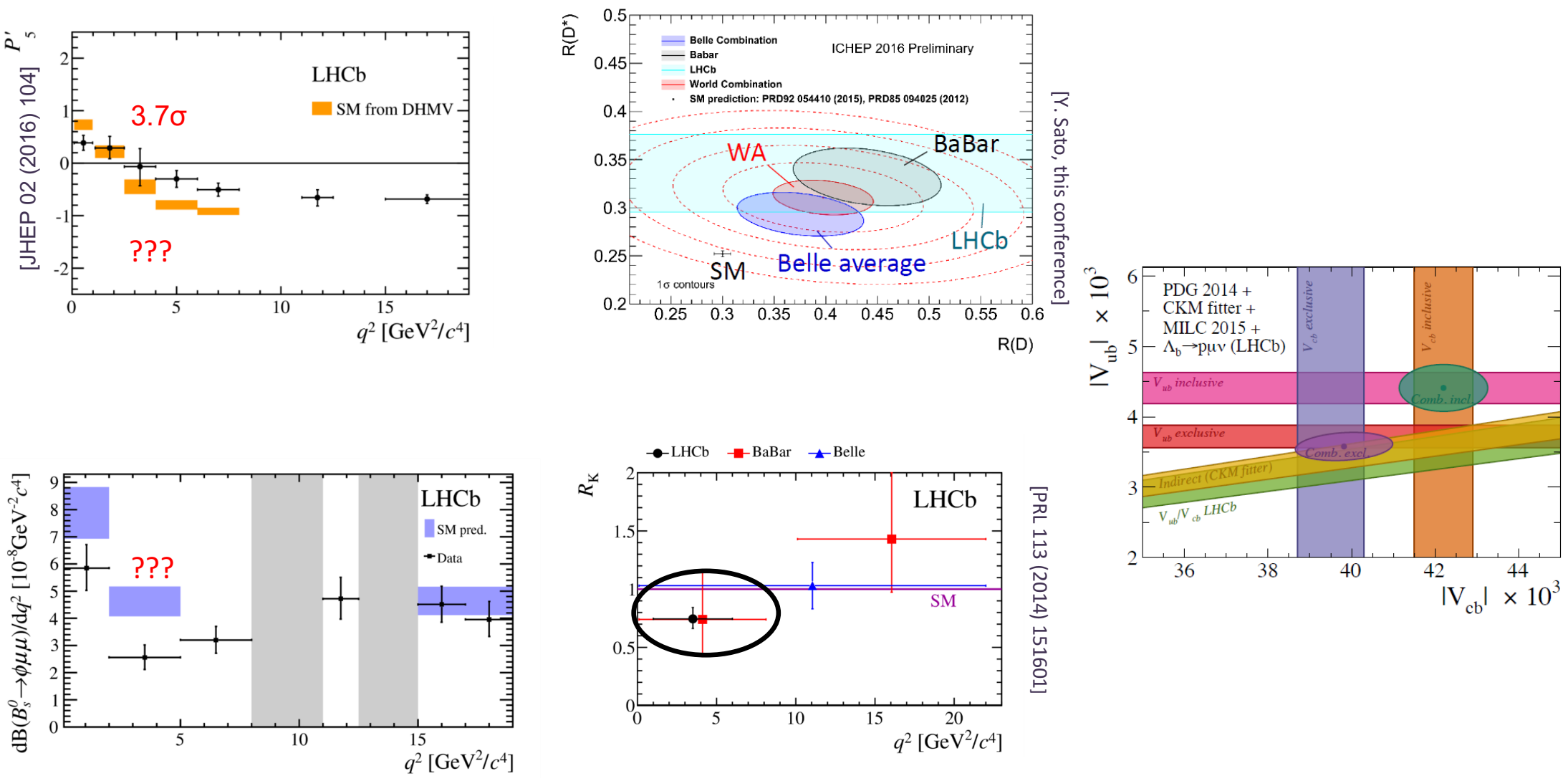

Figure 12: LHCb data show anomalous behaviour in $b \rightarrow sl^+l^-$ observables: (top left) $B^0 \rightarrow K^*\mu\mu$ in the $P_5$' - $q^2$ distribution and (bottom left) in the differential rate of $B_s \rightarrow \phi\mu\mu$. LHCb, BaBar and Belle data indicate hints of lepton universality violation in: (top middle) $B \rightarrow D^{(*)}l\nu$ and (bottom middle) in $B \rightarrow Kl^+l^-$. (Right) Longstanding inconsistency in exclusive vs. inclusive $V_{ub}$ and $V_{cb}$ determinations in date from a variety of experiments including LHCb, BaBar, Belle and CLEO.

There are a large range of smaller and tabletop experiments searching for fundamental physics, made possible by new technologies and theoretical ideas that permit new ways to probe the Universe. Some examples include:

- Axion searches (CAPP, ADMX)
- Ultra precise EDM tests (electron, muons, nucleons)
- Precision Gravity (new forces) – Cold Atoms
- Towards Relic Neutrinos (e.g. Tritium)
- Probing Quantum Foam

Trapped ions are versatile: they enable precision spectroscopy (atomic clocks), comparing electric and magnetic transitions tracks variations in $\alpha$, can lead to tests of the weak equivalence principle, "gbar" to test the equivalence of gravitational acceleration (little g) between matter and antimatter. Also, optical clocks based on trapped ions (or neutral atoms) are sensitive to the gravitational redshift for height differences of only ~30 cm at sea level, enabling tests of general relativity. These small experiments involve new thinking and new expertise - they should be encouraged and supported.

Neutrinos are immensely important. Neutrinos are massive, constituting the only laboratory-based evidence for physics beyond the Standard Model. They are a gift and a window into the world of new physics. Neutrinos are the most copious particles in the Cosmos after photons, they delight poets, and they are favored by the Nobel Committee. They offer a rich flavor structure, connections to very high mass scales, and the possibility of leptogenesis. For all these reasons it is mandatory to study them in exquisite detail.

Neutrino Oscillations are a beautiful example of a quantum interference phenomenon with a baseline varying from meters up to the Earth-Sun distance in experiments conducted so far. These experiments indicate that the mass splittings between the different neutrino states is of order 0.03 eV or less. These fundamental particle mass differences are remarkably small, of the order of molecular excitation energies. The current long baseline program is making exceptional progress. While not yet conclusive the results presented at ICHEP show that neutrino physics is entering a new era of sensitivity and maturity.

Data from T2K currently favour the idea of CP violation in the lepton sector, which is one of the conditions required for the observed dominance of matter over antimatter in the universe, while data from NOvA disfavour the idea that mixing of the second and third neutrino flavours is maximal, representing a test of a new symmetry that underlies maximal mixing (Figure 13). Combining the T2K, Nova and SuperK experiments in a global fit, CP conservation is excluded at 2 sigma. This is the first robust indication of CP Violation in the leptonic sector, and the community is building the program to turn indication into observation.

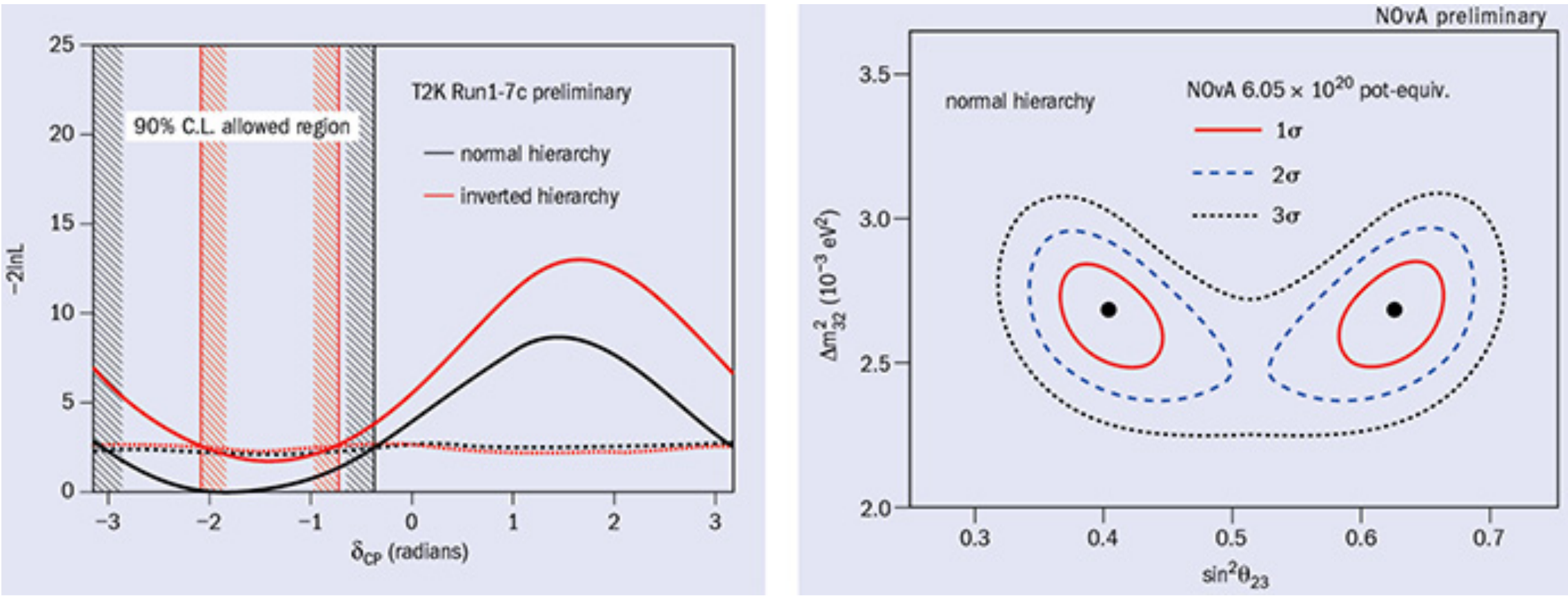

Figure 13: With nearly twice the antineutrino data in 2016 than in 2015 result, T2K observes an electron antineutrino appearance rate lower than would be expected if CP asymmetry is conserved (left). With data accumulated until May 2016, representing 16% of its planned total, the NOvA results (right) show a preference for non-maximal mixing – $\sin 2\theta_{23} \neq 0.5$.

Sterile neutrinos – hypothesised particles that do not interact via SM forces – also received new attention in Chicago. The 20 year-old signal from the LSND experiment at Los Alamos National Laboratory, indicates $4\sigma$ evidence for a sterile neutrino. As reported at ICHEP, however, cosmological data and new results from IceCube in Antarctica and MINOS+ at Fermilab do not confirm the existence of sterile neutrinos. Interestingly, the Daya Bay experiment in China, Reno in South Korea and Double Chooz in France confirm a reactor neutrino flux that is low compared with the latest modelling, which could arise from mixing with sterile neutrinos. However, all three experiments also confirm an excess in the neutrino spectrum at an energy of around 5 MeV, compared with predictions, shedding doubt on reactor flux modelling. There is a clear need for an influx of theorists to place nuclear matrix element calculations on a firmer footing. An extensive range of next generation sterile neutrino experiments are almost ready, both reactor neutrino experiments and an important short baseline neutrino oscillation program using the Fermilab Booster neutrino beam.

The quest to observe neutrinoless double beta decay, which is only possible if the neutrino is Majorana, continues. The existence of a Majorana neutrino will necessarily imply that lepton number is not a conserved quantity. That would be a tremendous discovery, comparable to the demonstration that parity was not conserved by the weak interaction in 1956. Current knowledge of the neutrino mixing parameters provides a firm prediction for the range of values of the parameter $m_{\beta\beta}$ in both hierarchies (NH favored). Tritium experiments have reached $m_{\nu e} <$ 2 eV, and expect to reach, with KATRIN, $m_{\nu e} <0.2$ eV. From cosmology: $\Sigma\, m_i < 0.23$ eV (95% CL) today, but in the next decade there are good prospects to reach, via multiple probes, a sensitivity at the level of $\Sigma\, m_i < 0.01$ eV. Therefore, it is timely and compelling to embark on a renewed discovery quest to observe neutrinoless double beta decay. These are incredibly difficult experiments, and some of the important experiments and their sensitivities were presented at ICHEP. To see how challenging: a half life $T_{1/2}$~$10^{26}$ years corresponds to $<m_\nu>$~50-100 meV. With 100kg of isotope one can expect ~1 event/yr! The next generation experiments will push into the inverted hierarchy region, and further into the future we can expect tonne scale experiments that will begin to reach the normal hierarchy with long exposure.

Here is what an intensity frontier discovery scenario might look like: neutrinoless double-beta decay is observed in the next generation of experiments. There are multiple confirmations with multiple nuclei. A major theory effort ensues to understand matrix elements leading to breakthrough methods and unprecedented accuracies (few %) on the mass. We are in the quasi-degenerate region, we have a lower limit on the mass of the lightest eigenstate. Then KATRIN observes this mass, a big version of the Project 8 experiment measures that mass very well. Cosmological measurements then find strong evidence for neutrino mass and measure the sum of the masses. Combining all data the Majorana CP phases must be non-zero! DUNE/Hyper-K come online and measure the Dirac phase delta with high precision. We are then able to home in on a leptogenesis model that explains the baryon asymmetry. What comes beyond Hyper-K and DUNE? The answer is very accurate measurements of the PMNS matrix at a neutrino factory. While this work in unfolding Super Belle/ LHCb produce very accurate CKM matrix elements determinations. So now we have very accurate data on flavor and on masses. Will that lead to an understanding of the fundamental origin of the CKM, PMNS, and mass matrices? We do not know, but the scenario described makes it plausible.

## 1.5 The cosmic frontier

From dark matter to dark energy and probing the Planck scale with inflation, there is a broad range of activities at the cosmic frontier that are marked by rapid, surprising and exciting developments: a wealth of particle physics using the whole universe as our laboratory.

Dark matter dominates the matter content of the universe, but its identity is still a mystery. Indeed, some theorists speculate about the existence of an entire "dark sector" made up of dark photons and multiple species of dark matter. Numerous approaches are being pursued to detect dark matter directly, and these are complemented by searches at the LHC, surveys of large-scale structure and attempts to observe high-energy particles from dark-matter annihilation or decay in or around our Galaxy. Regarding direct detection, experiments are advancing steadily in sensitivity: the latest examples reported at ICHEP came from LUX in the US and PandaX-II in China, and already they exclude a substantial fraction of the parameter space of supersymmetric dark-matter candidates (Figure 14, left). Future prospects include LZ, which should approach the "neutrino floor", at which time directional detection and other novelties will be needed (Figure 14, right).

Turning to indirect detection dark matter may pair annihilate or decay in our galactic neighborhood to positrons, high-energy photons, neutrinos, antiprotons, antideuterons… There are many fronts to cover. It is a tough field with astrophysical uncertainties playing an important role in the extraction of particle physics information. However we have a good record of learning about particle physics from the cosmos: examples include the discovery of **antimatter** (positron, Anderson, 1932), the discovery of the **second** generation (muon, Anderson, 1936), the discovery of the pion ("Yukawa" particles, 1947 Lattes, Powell, and Occhialini) and the demonstration of **neutrino mass and mixing** (1998-2001). Important recent/current experiments have been PAMELA, Fermi-LAT and AMS. Since 2010, electron and positron fluxes have been measured by AMS with remarkable precision up to ~400 GeV. Dark matter implications require precise determinations of cosmic ray fluxes, however. In indirect dark matter searches with photons, there have been rapid improvements in recent years. Fermi-LAT now excludes WIMPs with masses up to ~100 GeV, for certain annihilation channels. The future lies with the Cerenkov Telescope Array (CTA), which will extend the reach by two orders in mass, up to masses ~ 10 TeV.

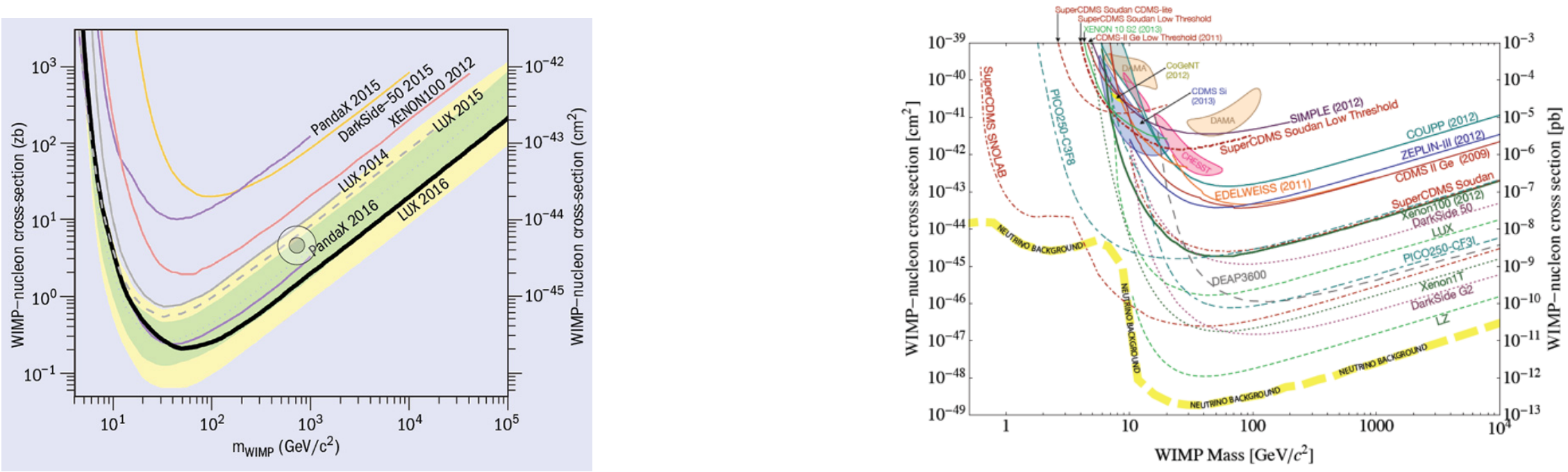

Figure 14: The LUX and PandaX experiments saw no signals of dark-matter candidates and their results were consistent with background expectations. A substantial region of parameter space is now excluded (left). Since 2010, sensitivity improved by ~100 (for m ~ 100 GeV). Further improvements by 2-3 orders of magnitude expected by a suite of experiments world-wide (right).

Here is what a dark matter discovery timeline might look like for a two component dark matter made up of WIMPS and axions. The scenario has been developed by Tim Tait:
2017: Xenon 1T sees a handful of events consistent with a dark matter mass < 400 GeV
2017: HESS observes a faint gamma ray line coming from the galatic center
2018: Two LHC experiments see a signifcant excess of leptons with missing energy
2018: SuperCDMS sees a similar signal to Xenon 1T
2019: Neutrinos are seen coming from the sun by IceCube
2019: Two LHC experiments do not see a significant excess of jets and missing energy
2020: A possible signal of axion conversion is seen in an upgraded ADMX
2030?: Observation at a Higgs factory indicates the cross-section to interact with leptons is too large to satisfy the relic density

Turning to cosmic surveys, inflation at early times shapes the CMB at 300,000 years after the big bang when matter and photons decouple, and seeds structure formation. The latter is driven by dark matter producing the growth of struture which is in turn driven by dark energy at late times, and by neutrinos which have a significant impact on the growth of structure at small scales.

Cosmic Surveys have been engines for science over the past several decades, as the era of precision cosmology has dawned. The CMB surveys have grown in scope from satellites like WMAP to Planck and, in the future, the proposed LiteBIRD. Ground based CMB surveys will find their most mature expression in the proposed CMB Stage IV mission. Optical surveys have likewise grown in scope from SDSS to DES to LSST, now under construction, and DESI, and the space-based Euclid and proposed WFIRST. As the survey field matures detailed comparisons with much richer data sets, using multiple complementary probes, will directly address the physics of inflation and dark energy. Information on dark energy can be gleaned using both optical surveys of large-scale structure and data from surveying the cosmic microwave background.

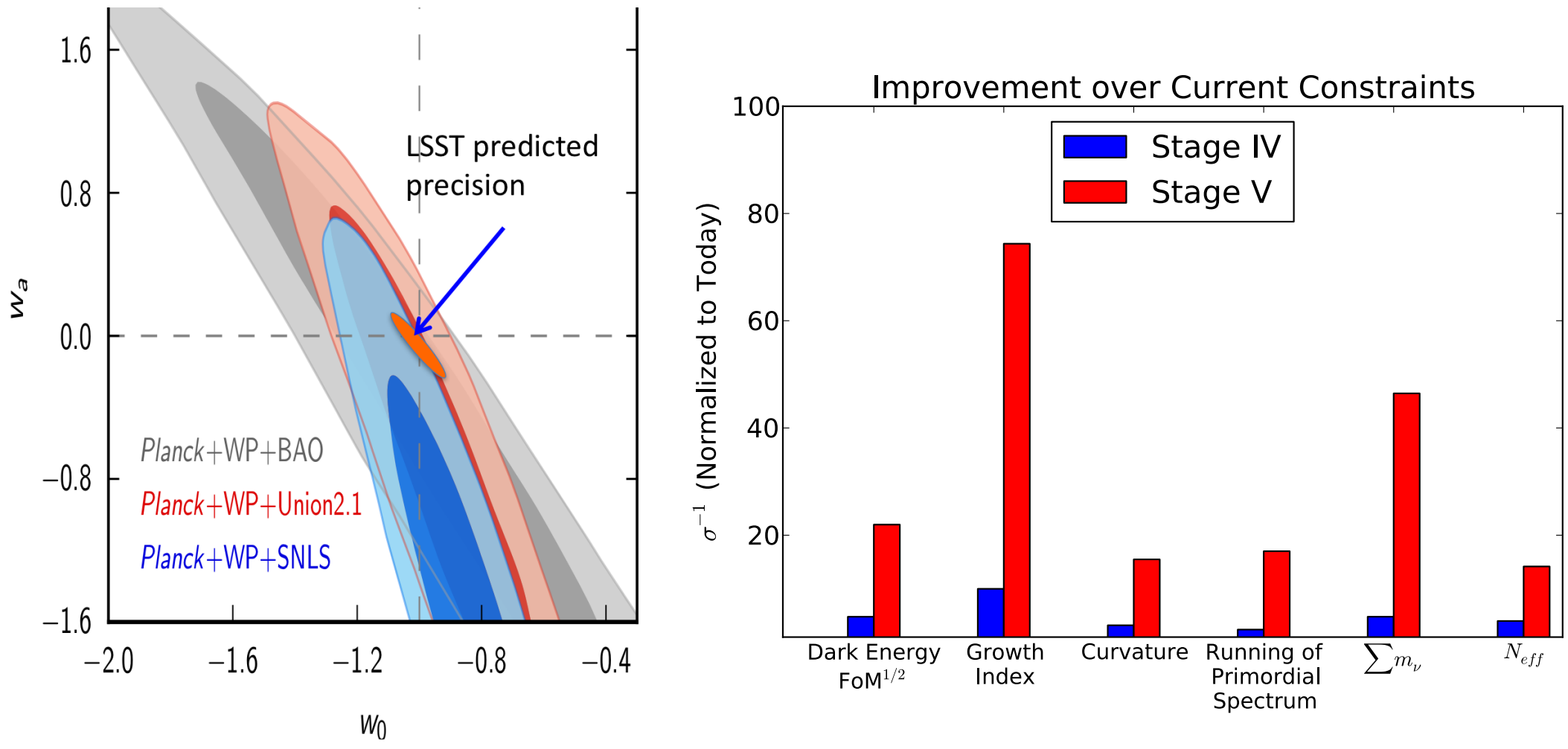

Figure 15 : (left) current knowledge (grey, blue and salmon colored ellipses) of the dark energy parameter $w_0$ (the ratio of pressure to the energy density) compared to $w_a$ the rate of change with time of dark energy derived from measurements of Type 1a supernovae and baryon acoustic oscillations in the spatial distribution of galaxies. The predicted precision of LSST using multiple probes including the measurement of cosmic shear via weak lensing is the small orange ellipse. (right) stage V and its comparison to stage IV is shown here in terms of inverse precision normalized to today for dark energy, the sum of the neutrino masses and other quantities.

The are numerous probes of dark energy: the distribution of objects of known brightness in the sky, the angular diameter of objects of known size as a function of distance, structure growth in the universe as a function of redshift or time by counting clusters or through the measurment of weak lensing. Figure 15 shows the the impact of LSST (a stage IV dark energy mission) compared to the current knowledge of dark energy (which is known to about 10%) and the rate of change of dark energy with time where current knowledge does not constrain it. LSST makes an important contribution to measuring dark energy and is complementary to other future missions: ex: Euclid/DESI etc. all of these missions are needed. Even after the stage IV missions have met their obejectives, there will still be much more information to extract from the sky.

Cosmic Visions, a community activity of US particle physicists and cosmologists, have nicely summarised this in a recent report from which Figure 15 (right) is taken.

Here is one hypothetical example of what a discovery scenario might look like at the cosmic frontier:

2018: DES finds hints of a large neutrino mass sum

2020: Stage 3 CMB experiments find hints of B-modes

2023: Neutrino-less double-beta decay detects Majorana neutrinos

2025: CMB-S4, DESI/LSST/Euclid measure neutrino masses to 6-sigma

Neutrino mass structure, Seesaw scale of $10^{15}$ GeV

2026: LSST/DESI/Euclid find hints of primordial non-gaussianity (PNG)

2028: CMB-S4 confirms the tensor/scalar ratio $r = 0.05$ at $20\sigma$; inflation scale is $10^{16}$ GeV

2035: DUNE/Hyper-K discovers proton decay

2045: 21 cm experiments detect a vast range of types of PNG, constraining the effective Lagrangian that drove inflation. This leads to a full Grand Unified Theory with confirmed predictions for inflation, neutrinos, and proton decay.

### 1.6 Instrumentation the Great Enabler

Instrumentation is the great enabler. *"New directions in science are launched by new tools much more often than by new concepts. The effect of a concept-driven revolution is to explain old things in new ways. The effect of a tool-driven revolution is to discover new things that have to be explained"* – Freeman Dyson. In our field, we detect & measure over 24 orders of magnitude in energy, from the CMSB to cosmic rays. We use a rich spectrum of technologies (Figure 16).

Our instrumentation represents both a towering achievement, and, in some cases, a scaled-up version of techniques used in the past. Many experiments are large and have high costs resulting in major de-scoping of detectors and their capabilities, to the detriment of physics reach, to match available resources. Instrumentation R&D has the power to transform this situation. We need to develop new technologies to find new physics. A multi-disciplinary environment is needed where particle physics partners with other disciplines involving both academia and industry, and this must be enhanced and strengthened. We had a well-attended session at ICHEP on Technology Applications and Industrial Opportunities, with speakers from academia, laboratories, industry and intergovernmental institutions. Presentations covered: innovation, strategies for research laboratories, applications (medicine, aerospace, material science etc.), entrepreneurship and start-up, success stories and industry perspectives. What are the best models to bring developments from HEP to the market? How can we lower the current barriers to realizing industrial opportunities? HEP as mediator? Industry as mediator? We need proof of concept support for maturing technologies. There is a need for international collaborations and networks, and help to young people from our field, many of whom go into industry, to develop as entrepreneurs creating spin offs and start-ups. Industry is looking for talent from our community- let's help them find it.

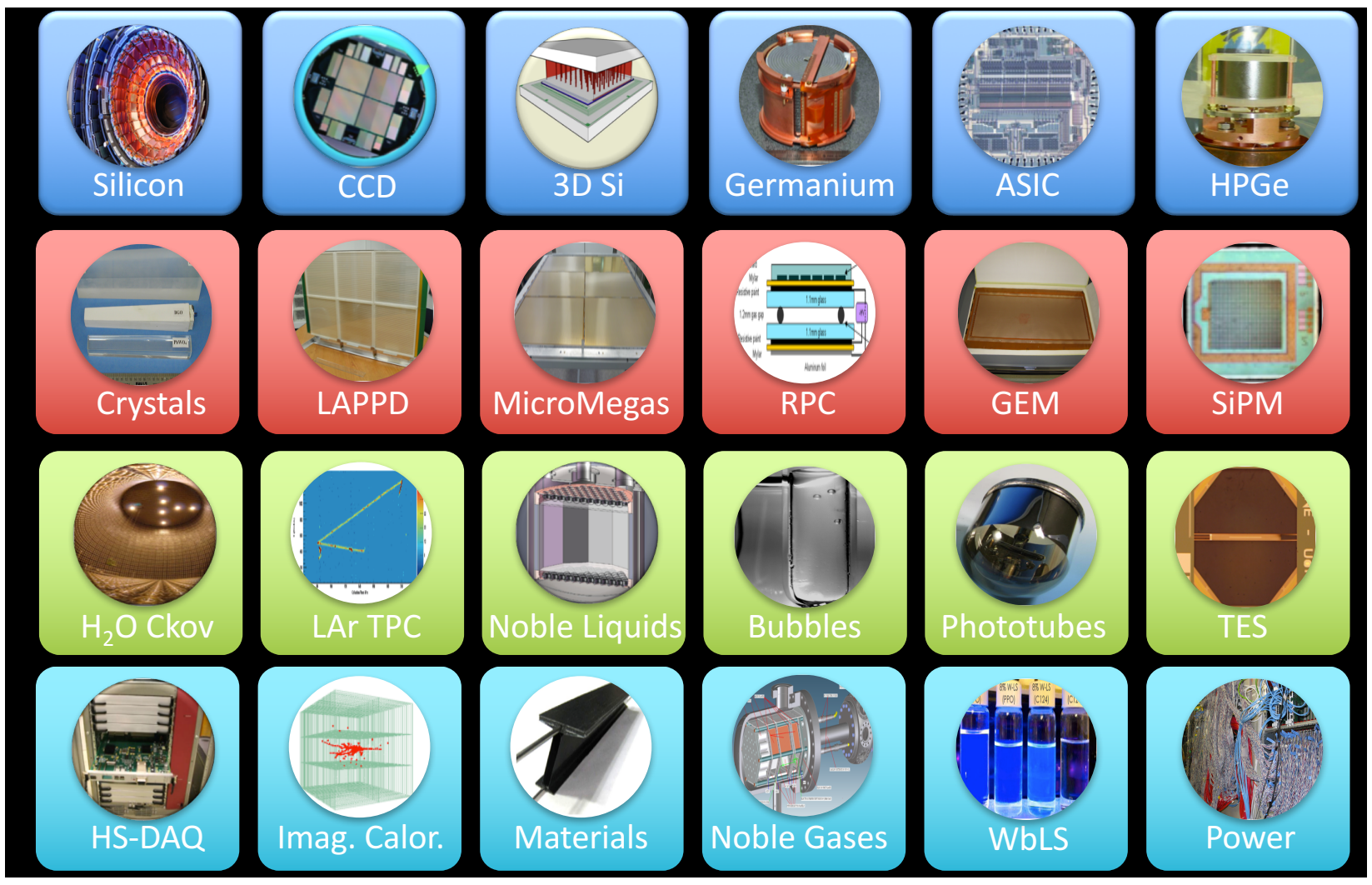

Figure 16: A selection of the detection technologies used in particle physics. (Original figure from Marcel Demarteau.)

There is good reason to be optimistic that instrumentation will continue to be a great enabler of science. This is in part due to the spectacular gains of the microelectronic and telecommunications industries from transistor count and ADC power reduction, to data storage and bandwidth gains. Doubling times of less than two years have been the norm up to now. Certainly some scalings will stop, but new approaches will come on line. One of the most promising areas is CMOS sensors for tracking and calorimetry, offering low cost, low mass, potentially rad hard sensors for high energy pp and ee colliders and the intensity frontier. In the area of Trigger & DAQ: R&D in Associative Memories, FPGAs, GPUs, CPUs, Communications Industry Architectures and Link Technologies are all promising. Today at the LHC the increased luminosity means increased pileup, which is exacerbated because we measure in three dimensions. But if we opened the fourth dimension through precision timing, through a combination of silicon and calorimetry, we can measure the time of neutral energy in the event with a resolution of 30 ps – a potentially transformative situation. Trigger and DAQ are likely to be transformed as well. Developments in exascale heterogeneous and neuromorphic computing, and powerful and flexible intelligent trigger tools will enable low thresholds triggers to exploit increased instantaneous luminosity to maximise physics, and machine learning will be ubiquitous.

Quantum sensors enable the study of the early universe and the search for dark matter. Transition edge sensors and kinetic inductance detectors have broad applications at the cosmic and intensity frontiers and in X-ray spectroscopy and photon science. Photon detection is critical and ubiquitous over a wide range of wavelengths and signal times. The development of large-area devices that are radio-pure, with cryogenic stability and high quantum efficiency within an appropriate wavelength sensitive window would be a "game-changer" with significant impact in areas outside of high energy physics.

Moving to the seemingly mundane question of cables: wireless technology is ubiquitous and, for most of the world, indispensable. In contrast, in particle physics cable plants are (still) ubiquitous. Can we use wireless instead of cables? Richard Brenner at Upsala has built a 60 GHz wireless readout system with the following features: 4.5Gbps @1m, 240mW power consumption, and a Bit Error Rate < $4\text{x}10^{-15}$; this is a promising start. Additive manufacturing has ushered in a new era of opportunity as well. There are techniques being developed such as 3D printing that may be important for particle physics, but we have not learnt how to use them yet - but you can make a car already (a 3D Shelby Cobra) and it drives, and has been driven by U.S. Secretary of Energy Ernie Moniz.

## 1.7 Theorists build cathedrals too!

The ATLAS and CMS detectors are modern day cathedrals, but particle theory has been building cathedrals too. Theory is advancing rapidly along two main lines: new ideas and approaches for dark matter and naturalness, and more precise calculations of SM processes that are relevant for on-going experiments. As emphasised at ICHEP 2016, new ideas for the identity of dark matter have had implications for LHC searches and for attempts to observe astrophysical dark-matter annihilation, in addition to motivating a new experimental programme looking for dark photons.

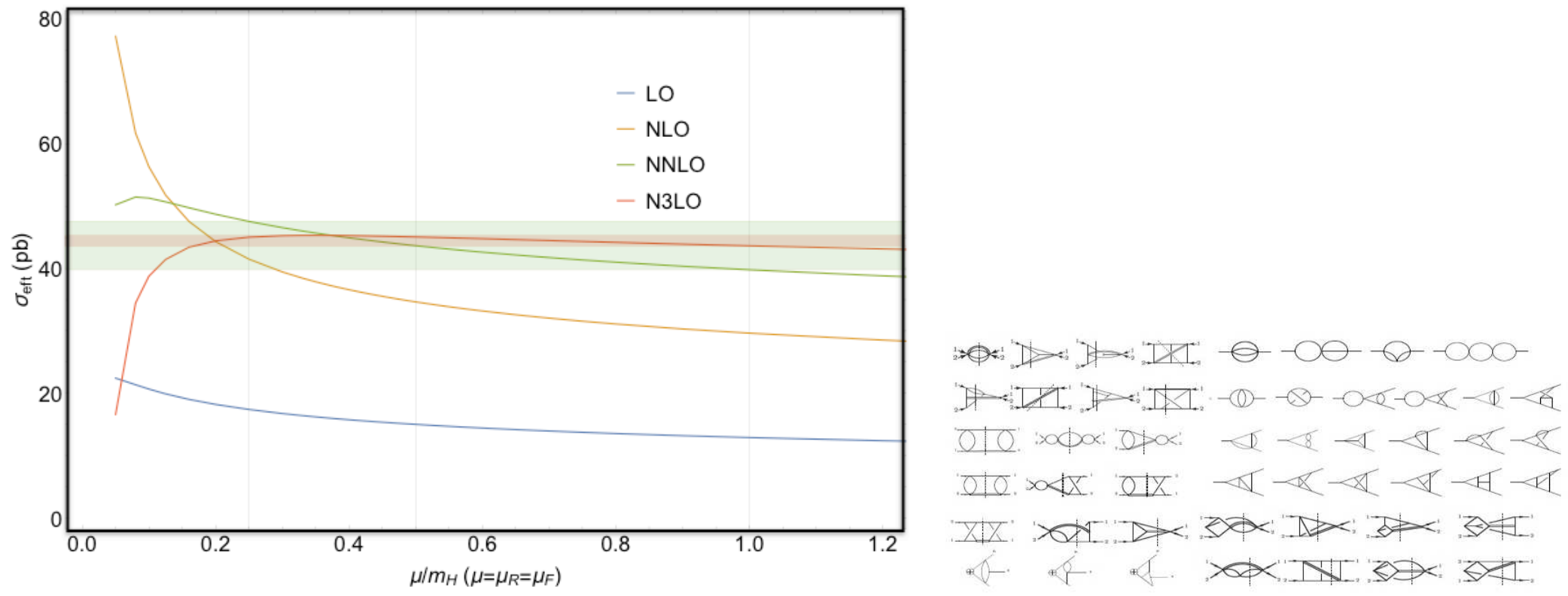

Figure 17: Calculating the Higgs cross section at N3LO has been a remarkable achievement (left), some of the diagrams involved (right).

There have also been tremendous developments in theoretical calculations with higher-order QCD and electroweak corrections (see Figure 17), which are critical for understanding the SM backgrounds when searching for new physics – particularly at the LHC and, soon, at the SuperKEKB B factory in Japan. The LHC's experimental precision on top-quark production is now reaching the point where theory requires next-to-next-to-next-to-leading-order corrections, and these are starting to be produced. In addition, recent lattice QCD calculations play a key role in extracting fundamental parameters such as the CKM mixing matrix elements, as well as reducing uncertainties to where effects of new phenomena beyond the SM may conclusively emerge.

### 1.8 Accelerator Frontier

Particle physics is a global endeavor; the LHC at CERN serves as the exemplar of a successful large international science project. At a session devoted to future facilities, the science case and current status of new projects that require international co-operation was presented. The International Linear Collider (ILC) in Japan is shovel ready and awaiting a decision from MEXT; the Circular Electron–Positron Collider (CEPC) in China is now the number one priority of the Chinese particle physics community, which is an important step towards realisation; an energy upgrade of the LHC is a possible stepping stone to FCC-hh; the Compact Linear Collider (CLIC) foresees realisation in 3 stages over a 20-30 year horizon, starting with a Higgs/top factory at around 380 GeV; the Future Circular Collider (FCC) at CERN; the Long-Baseline Neutrino Facility (LBNF) and Deep Underground Neutrino Experiment (DUNE) in the US; and the Hyper-K neutrino experiment in Japan.

Many ideas are being developed for multi-TeV colliders with gradients >100 MeV/m, and with lower capital and operating costs: Wakefield Acceleration using plasmas or dielectrics and direct laser acceleration. Both particle beam (PWFA) and laser (LWFA) driven wakefield approaches are thought to offer effective gradients of O(1 GeV/m). Increased emphasis and support should be provided to train the next generation of accelerator scientists. This is necessary for our community to be able realise the aspirations of building next generation machines – ILC or CLIC, FCC, CEPC or HE-LHC, or a plasma-based collider.

Enabling machines like these to become reality will require an immense amount of political support and funding as well. It will require advances in accelerator R&D to make the FCC and variants affordable and advances in detector R&D to lower the cost of detectors. It is critical that, at the same time as we plan the large-scale facilities, we maintain space in our program for experiments at small and medium scale - they play a crtiical role in our field and its progress. It is important that the audience today, and in fact the entire particle physics community, become or remain very engaged in the national and regional physics communities, and in the decision-making process to ensure a bright future.

### 1.9 Celebrate our diversity

One of the well-attended sessions at ICHEP 2016 concerned professional issues critical to a successful future for the field of particle physics. Diversity and inclusion were the subject of parallel sessions, discussions and posters, with themes such as communication, inclusion and respect in international collaboration, and how harassment and discrimination in scientific communities create barriers to access. The sessions were mostly standing-room only, with supportive but candid discussion of the deep divides, harassment, and biases – both explicit and implicit – that need to be overcome in the science community. Speakers described a number of positive initiatives, including the Early Career, Gender and Diversity office established by the LHCb collaboration, the Study Group on Diversity in the ATLAS collaboration, and the American Physical Society's "Bridge Program" to increase the number of physics PhDs among students from under-represented backgrounds [4].

**2.0 Summary**

We must believe our science is compelling enough to compete favorably for the best talent in a world where transformational and paradigm-altering advances are happening in other fields such as biology and energy research. It is a privilege to receive public funding and we must strive to be worthy of it. The compelling opportunities that we want to pursue must compete favorably with other opportunities on all the playing fields: in the agencies, in Congress, and in academia.

We must explain why particle physics is important to pursue to governments, and to our colleagues in the academy who are not particle physicists, and may not be physicists at all. To do this we must see our science as compelling, and we need to convince others it is compelling as well. A healthy program needs a long-term strategy with a compelling vision for the future and future scientific achievements. This is what our field has produced.

Our community have identified compelling opportunities at all scales. We must participate fully in the national and global planning processes. Continue to reach out to our community; especially to younger colleagues, give them roles; they are the future. Maintain the high level of interaction between the different parts of our community across all of our specialties and all of our regions. We are one field with one voice. The leadership of the community must endeavor to continue to develop structures that will maintain the sense of community and global coordination that will be needed to achieve our aspirations

There is a vast array of scientific opportunities now and in the future with which to further explore the smallest and largest structures in the universe. The LHC is performing beyond expectations, neutrino physics has progressed dramatically, and its progress will continue, intense kaon and muon beams, and SuperKEKB, and an ambitious program to probe the nature of dark matter and dark energy and to further study the CMB provide a variety of ways to look at the universe on all scales. There has never been a more exciting time to be a particle physicist!

Finally, I would like to thank, on behalf of all of the speakers and attendees at ICHEP 2016, Young Kee Kim and her team for a remarkable and memorable conference in the great City of Chicago! What we know is a droplet what we don't know is an ocean. The ocean is for you to explore. An update on what you have found in the ocean will be given in two years at ICHEP 2018 in Seoul.

## Acknowledgment

In preparing for this talk I benefitted from conversations with and material from: Artur Apreysan, Phil Allport, Bill Barletta, Alan Barr, Steve Biller, Ed Blucher, Oliver Buchmueller, Phil Burrows, Daniela Bortoletto, Themis Bowcock, Joel Butler, Jim Brau, Bill Brinkman, Chip Brock, Rick Cavanaugh, Janet Conrad, Chris Damerell, Persis Drell, Andre De Gouvea, Marcel

Demarteau, Jonathan Feng, Brian Foster, Fabiola Gianotti, Marat Gautaulin, Tim Gershon, Saul Gonzalez, Howard Gordon, Al Goshaw, Nick Hadley, Joanne Hewett, Joe Incandela, Steve Kahn, Young-Kee Kim, Joe Lykken, Boaz Klima, Rocky Kolb, Jesse Liu, Hitoshi Murayama, Dan Marlow, Michelangelo Mangano, Felicitas Pauss, Moishe Pripstein, Andrei Seryi, Yves Sirois, Jo Spitz, Tim Tait, Mark Trodden, Mike Tutts, Tony Tyson, Lucianne Walkowicz, Dave Wark, Guy Wilkinson, Katie Yurkewicz and many more. I thank Ben Jones for editorial assisistance.